\documentclass[prd
,preprint
,superscriptaddress,showpacs,nofootinbib,%
tightenlines]{revtex4}
\usepackage{epsfig}
\usepackage{amssymb,amsmath}
\usepackage{ulem}
\newcommand{\beq}{\begin{equation}}
\newcommand{\eeq}{\end{equation}}
\newcommand{\beqa}{\begin{eqnarray}}
\newcommand{\eeqa}{\end{eqnarray}}

\newcommand{\ben}{\begin{displaymath}}
\newcommand{\een}{\end{displaymath}}
\newcommand{\be}{\begin{equation}}
\newcommand{\ee}{\end{equation}}
\newcommand{\bea}{\begin{eqnarray}}
\newcommand{\eea}{\end{eqnarray}}
\begin{document}
\title{
$^1S_0$ nucleon-nucleon scattering in the modified Weinberg approach}
\author{E.~Epelbaum}
\affiliation
{Institut f\" ur Theoretische Physik II, Fakult\" at f\" ur Physik und Astronomie, \\ Ruhr-Universit\" at Bochum 44780 Bochum, Germany}
\author{A.~M.~Gasparyan}
\affiliation
{Institut f\" ur Theoretische Physik II, Fakult\" at f\" ur Physik und Astronomie, \\ Ruhr-Universit\" at Bochum 44780 Bochum, Germany}
\affiliation
{SSC RF ITEP, Bolshaya Cheremushkinskaya 25, 117218 Moscow, Russia}
\author{J.~Gegelia}
\affiliation
{Institut f\" ur Theoretische Physik II, Fakult\" at f\" ur Physik und Astronomie, \\ Ruhr-Universit\" at Bochum 44780 Bochum, Germany}
\affiliation
{Tbilisi State University, 0186 Tbilisi, Georgia}
\author{H.~Krebs}
\affiliation
{Institut f\" ur Theoretische Physik II, Fakult\" at f\" ur Physik und Astronomie, \\ Ruhr-Universit\" at Bochum 44780 Bochum, Germany}

 \date{12 July, 2014}

\bigskip
\begin{abstract}
\noindent
Nucleon-nucleon scattering in the  $^1S_0$ partial wave is considered  in
chiral effective field theory within the renormalizable formulation of
Ref.~\cite{Epelbaum:2012ua} beyond the leading-order approximation. By applying
subtractive renormalization, the subleading contact
interaction in this channel is taken into account non-perturbatively. For a
proper choice of renormalization conditions, the
predicted energy dependence of the phase shift and the
coefficients in the effective range expansion are found to be in a good agreement
with the results of the Nijmegen partial wave analysis.
\end{abstract}
\maketitle

\section{Introduction}
\label{sec1}

The seminal work of Weinberg \cite{Weinberg:rz} has triggered a
renewed interest to the nuclear force problem in the framework of effective field theory (EFT).
In this approach, nuclear forces are defined as kernels of the
corresponding dynamical equations and can be derived order-by-order
making use of the systematic chiral expansion.

Starting from the pioneering work of Ref.~\cite{Ordonez:1995rz}, this
approach has developed rapidly and is nowadays commonly
employed in studies of low-energy few- and many-nucleon dynamics and
nuclear structure calculations, see
\cite{Epelbaum:2008ga,Machleidt:2011zz,Epelbaum:2012vx} for recent
review articles.
While offering many attractive features, Weinberg's approach was
criticized for being non-renormalizable.
The main difficulty is related to the fact that iterations of
the truncated NN potential within the Lippmann-Schwinger (LS) equation generate
ultraviolet (UV) divergencies which cannot be absorbed by counter terms (contact
interactions) included in the truncated potential.
In particular, infinitely many counter terms are needed to absorb
UV divergences emerging from iterations of the leading-order (LO) one-pion exchange (OPE)
potential \cite{Savage:1998vh}.
This feature is sometimes referred to as inconsistency of Weinberg's approach.

The UV cutoff  $\Lambda$ can be removed from the LS
equation by enforcing the limit $\Lambda \to \infty$ non-perturbatively, see
e.g.~\cite{Nogga:2005hy,PavonValderrama:2005gu}. It is possible to obtain a
finite, manifestly non-perturbative solution of the LS equation with
a singular $1/r^3$-potential by including one/no contact operator in
each attractive/repulsive channel \cite{PavonValderrama:2005gu}. However,
such a procedure is incompatible with the principles of EFT which
require that all UV divergences emerging from iterations of the LS equation
are absorbed by counter terms \cite{Epelbaum:2009sd}. It is not
surprising that such an approach fails to reproduce experimental data even at N$^3$LO \cite{Zeoli:2012bi}.

Treating the exchange of pions perturbatively as suggested
by Kaplan, Savage and Wise (KSW) \cite{Kaplan:1998tg} allows one
to avoid the above-mentioned inconsistency. However, the perturbative series fail to converge within this
framework  \cite{Gegelia:1998ee,Cohen:1998jr,Gegelia:1999ja,Fleming:1999ee}.

Presently, there exist different views and formulations of organizing the chiral expansion in
the few-nucleon sector
\cite{Nogga:2005hy,PavonValderrama:2005gu,Epelbaum:2009sd,Lepage:1997cs,Gegelia:1998iu,Gegelia:2004pz,Epelbaum:2006pt,Mondejar:2006yu,
Long:2007vp,Yang:2009pn,Birse:2010fj,Valderrama:2009ei,Valderrama:2011mv,Long:2011xw,Long:2012ve,Long:2013cya,Beane:2008bt,Gasparyan:2012km,Gasparyan:2013ota}.
A novel approach to the NN scattering problem in EFT has been formulated in
Refs.~\cite{Epelbaum:2012ua,Epelbaum:2012cv,Epelbaum:2013ij,Epelbaum:2013fwa,Epelbaum:2013naa}
and is referred to as the modified Weinberg approach.
Within this framework, the leading order (LO) NN scattering amplitude
is obtained by solving the Kadyshevsky equation \cite{Kadyshevsky:1967rs}
for the LO potential consisting of the contact interaction part and
the OPE potential. This equation provides an  example of
three-dimensional integral equations which satisfy relativistic
elastic unitarity.
An important feature of the Kadyshevsky equation is that it is
renormalizable for the LO potential, i.e.~all ultraviolet divergences
generated by iterations can be explicitly absorbed into redefinition
of the NN derivative-less contact interaction. The scattering amplitude can
still be renormalized if higher-order corrections to the
potential are taken into account perturbatively.
If higher order corrections to the potential indeed provide small
contributions to the amplitude,
their perturbative and non-perturbative inclusions are expected to lead to small
differences in the results which are beyond the accuracy one is
working at. However, this observation is only meaningful if a proper
renormalization is carried out in both cases. In general, we are not
able to subtract all divergences from amplitudes if higher-order
contributions in the potential are treated non-perturbatively. In
  the $^1S_0$ partial wave, one observes a very large discrepancy between the LO EFT results and the
experimental data already at rather low energies
\cite{Epelbaum:2012ua}.
This large discrepancy signals that at least a part of the higher-order contributions in the effective
potential is likely to require a non-perturbative
treatment within our approach.\footnote{Notice, however, that the LO
  calculations reported in
  Refs.~\cite{Epelbaum:2014efa,Epelbaum:2014sza} within the standard nonrelativistic approach using a
  coordinate-space regularization for the OPE potential yields
  a superior description of the phase shift.
  }

In this paper we study in detail the role of the next-to-leading order (NLO)
short-range contribution to the potential which can be included both
perturbatively and non-perturbatively.
Specifically, we will
express the solution to the integral equation in a closed form
following the lines of Ref.~\cite{Gegelia:2001ev} and
apply the BPHZ-type subtractive renormalization
\cite{Collins:1984xc}. After subtracting {\it all} ultraviolet
divergences, we will calculate the remaining finite expressions
numerically, fit the available two low-energy constants (LECs) to the data and
compare the obtained results with the phase shifts for various choices
of the renormalization scale parameter. Here and in what follows,  the
resulting NN amplitude will be referred to as NLO as opposed to the LO result of
Ref.~\cite{Epelbaum:2012ua}.  A more complete calculation including
the corresponding two-pion exchange potential to first order in
perturbation theory is postponed for a future study.

Our paper is organized as follows: In section \ref{sec2} we provide
the formal expression for the scattering amplitude by making use of
the standard two-potential formalism. Subtractive renormalization of
the amplitude is discussed in detail in section \ref{sec2a}.
Next, section \ref{sec3} addresses the issue of the appropriate choice of the
renormalization conditions (i.e. subtraction scale) and also presents
the results of our calculation. Our findings are summarized in section
\ref{sec4}.

\section{Formal expression for the scattering amplitude}
\setcounter{equation}{0}
\def\theequation{\arabic{section}.\arabic{equation}}
\label{sec2}

In the framework of the modified Weinberg approach, the NLO $^1S_0$
partial wave NN scattering amplitude\footnote{Note the different
  overall sign
in comparison with the Feynman amplitude considered in Ref.~\cite{Epelbaum:2012ua}.} can be obtained by
extracting the $S$-wave component from the solution to the integral
equation (for the fully off-shell amplitude $T$)
\begin{eqnarray}
T\left(p_0,
\vec p\,',\vec p \, \right)&=&V \left(
\vec p\,',\vec p \, \right) + \int d^3 k \;
 V  \left(
\vec p\,',\vec k \, \right)G(p_0,k) \ T  \left(p_0,
\vec k,\vec p\right)
\,,\label{MeqLOk0integrated}\\
G(p_0,k)&=&\frac{m^2}{2(2\,\pi)^3}\frac{1}{\left(\vec k^2+m^2\right)\left(
p_0-\sqrt{\vec k^2+m^2}+i \epsilon\right)}\,,
\label{G}
\end{eqnarray}
where $\vec p$ ($\vec p \, '$) is the incoming (outgoing) three-momentum of
the nucleon in the center-of-mass frame, $p_0 = \sqrt{\vec q \, ^2 +
  m^2}$ with $m$ denoting the nucleon mass and $\vec q$ being the
corresponding three-momentum of an incoming (on-mass-shell) nucleon. Further, the potential is given by
\bea
\label{LO}
V\left(
\vec p\,',\vec p \, \right) & = & \left[C + C_2 \left(
\vec p\,'^2+\vec p\,^2 \, \right)\right] - \frac{g_A^2 M_\pi^2}{4
F_\pi^2}   \frac{1}{\left(\vec p\, '-\vec p \, \right)^2 + M_\pi^2}
\nonumber \\
&\equiv& V_C+V_\pi,\nonumber\\
C & = & C_S-3\,C_T+\frac{g_A^2}{4\,F_\pi^2}+D\,M_\pi^2.
\eea
Here $g_A$, $F_\pi$ and $M_\pi$ are the nucleon axial-vector coupling,
pion decay constant and the pion mass, respectively. The parameters $C_S$,
$C_T$, $C_2$ and $D$ refer to the LECs  of the effective
Lagrangian. Below, we work with the $S$-wave component of
Eq.~(\ref{MeqLOk0integrated}) and denote the $^1$S$_0$ partial wave projected OPE potential
by $V_\pi(p',p)$ with
\begin{eqnarray}
 V_\pi(p',p)=\frac{g_A^2 M_\pi^2}{16F_\pi^2 p p'}\ln\frac{(p-p')^2+M_\pi^2}{(p+p')^2+M_\pi^2}\,.
\end{eqnarray}
For the analysis of divergent integrals, it is useful to have the asymptotics of $V_\pi(p',p)$ at large values of momenta
\begin{eqnarray}
V_\pi(p',p)\big|_{\scriptscriptstyle p\to\infty,\,  p'<\infty}
  \approx-\frac{g_A^2 M_\pi^2}{4\,F_\pi^2\, p^2}\,, \quad \quad
V_\pi(p+l,p)\big|_{\scriptscriptstyle p\to\infty,\,|l|<\infty}\approx-\frac{g_A^2 M_\pi^2}{8\,F_\pi^2 \,p^2}\ln p\,.
\label{asymptotics}
\end{eqnarray}

The contact-interaction part of the potential $V_C$ is
separable. Therefore, it is possible to write the solution to
Eq.~(\ref{MeqLOk0integrated}) in a form, which allows one to carry out the subtractive renormalization explicitly.
This can be achieved by making use of the well-known two-potential
formalism. For this purpose, we write Eq.~(\ref{MeqLOk0integrated}) symbolically as
\begin{equation}
T=V+V\,G\,T,
\label{eqsim}
\end{equation}
and express its solution as
\begin{equation}
T=T_\pi+(1+T_\pi\,G)\,T_C (1+G\,T_\pi),
\label{Td}
\end{equation}
where $T_\pi$ and $T_C$ satisfy the equations
\begin{eqnarray}
T_\pi&=&V_\pi+V_\pi\,G\,T_\pi\,, \label{Tpi}\\
T_C&=&V_C+V_C\,G\,(1+T_\pi G)\,T_C\,.
\label{OEQ1}
\end{eqnarray}
For a separable contact-interaction potential,
\begin{equation}
V_C(p',p)= \xi(p')^T\, {\cal C}\xi(p),
\label{nuCfact0}
\end{equation}
where ${\cal C}$ and $\xi(p)$ are  $2\times 2$ and $2\times 1$ matrices, respectively, whose explicit form will be specified below,
the solution to Eq.~(\ref{OEQ1}) is also given in a separable form
\begin{equation}
T_C(p_0,p',p)= \xi^T (p') {\cal X}\xi(p),
\label{chiCfact}
\end{equation}
where ${\cal X}$  is a $2\times 2$ matrix,
\begin{eqnarray}
{\cal X}= \left[{\cal C}^{-1}-\Sigma\right]^{-1}\,,
\label{calX}
\end{eqnarray}
 and the $2\times 2$ ``selfenergy'' matrix ${\Sigma}$ reads
\begin{eqnarray}
\label{Sigma}
\Sigma(p_0)&=&\xi\,G\,\xi^T +  \xi\,G\,T_\pi G\,\xi^T \\
&\equiv &
\int d^3  k\;
  \xi(k ) G(p_0,k) \xi^T(k ) +
\int  d^3  k_1  d^3  k_2\;
  \xi(k_1 ) G(p_0,k_1) T_\pi   \left(p_0, k_1, k_2\right) G(p_0,k_2) \xi^T(k_2 )\,.
\nonumber
\end{eqnarray}
Thus, the final expression of the amplitude $T$ has the form
\begin{equation}
T=T_\pi+ \Xi^T\,{\cal X}\,\Xi.
\label{taup}
\end{equation}
with
\begin{eqnarray}
 \Xi(p_0,p)&=&\xi (1+G\,T_\pi) \equiv \xi(p)+\int d^3  k\,\xi(k )\,G(p_0,k) \, T_\pi   \left(p_0, k, p\right)\,.
\end{eqnarray}

\section{Renormalization of the scattering amplitude}
\setcounter{equation}{0}
\def\theequation{\arabic{section}.\arabic{equation}}
\label{sec2a}

The expression for the scattering amplitude in Eq.~(\ref{taup})
contains UV divergences.
We perform renormalization by applying the BPHZ procedure, i.e.~we
subtract all divergences and sub-divergences of the loop diagrams and
replace the LECs  with their renormalized, finite values. In
general, in renormalizable theories, subtractive renormalization can
be realized by counter terms in the Lagrangian. Chiral effective field
theory is renormalizable in the sense of effective field theories,
i.e.~all divergences can be absorbed into redefinition of an infinite
number of counter terms. To realize subtractive renormalization in the
considered problem, we would need to include the contributions of an
infinite number of counter terms of the effective Lagrangian.
Although this is possible for the case at hand by considering
energy-dependent counter terms, here we only show explicitly one
momentum- and energy-independent counter term $\delta z$ and
write the contact interaction potential in a separable form
\begin{equation}
V_C(p',p) =  C+C_2\left(p'^2+p^2\right)
=  \left(
                         \begin{array}{cc}
                           1, & p'^2 +\delta z\\
                         \end{array}
                       \right)
 \left(
 \begin{array}{cc}
 \tilde C &  C_2 \\
   C_2 & 0\\
 \end{array}
 \right)
\left(
  \begin{array}{c}
    1 \\
    p^2+\delta z \\
  \end{array}
\right).
\label{nuCfact}
\end{equation}
The new parameter is expressed as
\begin{equation}
\tilde C = C - 2\,C_2 \delta z.
\label{ccrel}
\end{equation}
Thus, the contact-interaction potential has the form
\begin{eqnarray}
{\cal C} = \left(
\begin{array}{cc}
\tilde C & C_2 \\
C_2 & 0 \\
\end{array}
\right),\ \ \
\xi(p) \equiv (\xi_1(p),\xi_2(p))^T=  \left(
                         \begin{array}{cc}
                           1, & p^2+\delta z \\
                         \end{array}
                       \right)^T.
\label{cxi}
\end{eqnarray}

\begin{figure}
\epsfig{file=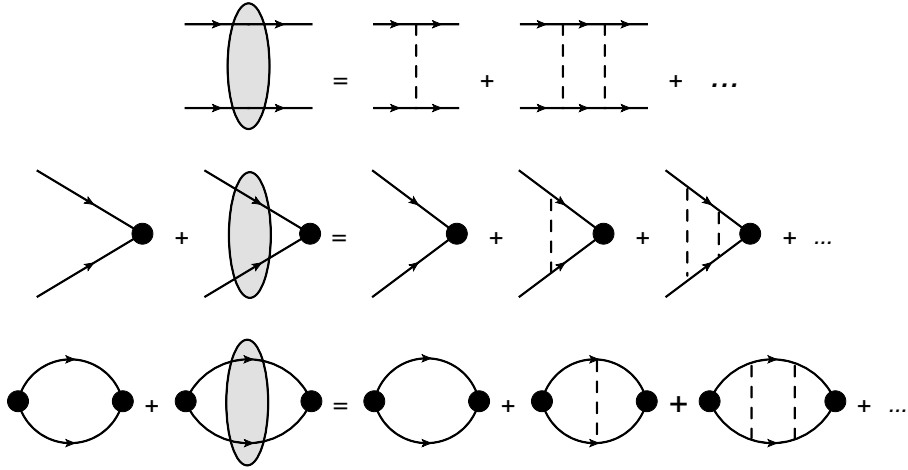, width=12cm}
\caption[]{\label{BB:fig} Building blocks of the scattering amplitude.
  The first, second and third lines represent  $T_\pi$,
 $\Xi$ and $\Sigma$, respectively. The solid and dashed
lines correspond to nucleons and pions, respectively. The filled
circles represent $\xi$ and $\xi^T$.
}
\end{figure}

The various terms contributing to the amplitude $T$ are visualized
diagrammatically in Fig.~\ref{BB:fig} in terms of the corresponding
building blocks, where
in the first line the amplitude $T_\pi$ is shown. The second line
represents $\Xi$, while the analogous diagrams for $\Xi^T$
are not shown explicitly. The third line depicts the quantity $\Sigma$
which contributes to ${\cal X}$, see Eq.~(\ref{calX}).
To obtain the amplitude using Eq.~(\ref{taup}), we first perform
subtractive renormalization and afterwards calculate numerically the
remaining finite expressions for the quantities $T_\pi$,
$\Xi$,  and ${\cal X}$.
In the following, we describe in detail how these quantities are
renormalized. Since
the amplitude $T_\pi$ is finite (the ultraviolet regularity of the equation for $T_\pi$~(\ref{Tpi}) follows from
 the asymptotics~(\ref{asymptotics})), we begin our discussion with the
subtractive renormalization of
$\Xi$.
By writing $\Xi(p_0,p)$ as a perturbative series as shown in Fig.~\ref{BB:fig},
\begin{equation}
\Xi = \xi+\xi \,G\, V_\pi+\xi \,G\, V_\pi\,G\,V_\pi+\cdots\,,
\label{barXipert}
\end{equation}
it is easily seen that $\Xi(p_0,p)=(\Xi_1(p_0,p),\Xi_2(p_0,p))^T$ can be obtained by solving the
integral equation
\begin{equation}
\Xi = \xi + \Xi \, G\, V_\pi\, .
\label{systemofeqs}
\end{equation}
This expression defines a system of equations for the quantities
$\Xi_{1,2}(p_0, p) $, which, using the
explicit form of $\xi(p)$ from Eq.~(\ref{cxi}), can be written as
\begin{eqnarray}
\Xi_1(p_0,p ) &=& 1+\int d^3  k\,\Xi_1(k)\, G(p_0,k) \, V_\pi(k, p)\, , \label{sofeq1x}\\
\Xi_2(p_0,p ) &=& p^2+\delta z+\int d^3  k\,\Xi_2(k)\, G(p_0,k) \, V_\pi(k, p)\,.
\label{systemofeqs2}
\end{eqnarray}
The equation for $\Xi_1(p_0,p)$ is free of ultraviolet divergences, see Eq.~(\ref{asymptotics}), and
has an ultraviolet behavior $\Xi_1(p_0,p)\stackrel{p\to\infty}{\approx} const$.
On the other hand, to identify the divergences in $\Xi_2(p_0,p)$, it
is convenient to consider iterations of Eq.~(\ref{systemofeqs2})
\begin{eqnarray}
\label{systemofeqs2it}
\Xi_2(p) &=& p^2+\delta z+\int d^3  k\,(k^2+\delta z)\,G(p_0,k)\,V_\pi(k, p) \nonumber\\
&+& \int d^3  k\,d^3  l\,(k^2+\delta z)\,G(p_0,k)\,V_\pi(k,l)\,G(p_0,l) V_\pi(l, p)+\cdots .
\end{eqnarray}
Remembering the definition of $G(p_0,k)$ in Eq.~(\ref{G}), we simplify
\begin{eqnarray}
k^2 G(p_0,k) &=& q^2 G(p_0,k) - \frac{m^2}
{2(2\,\pi)^3}\frac{p_0+\sqrt{\vec k^2+m^2}}{\vec k^2+m^2}
\nonumber \\
&\equiv & q^2 G(p_0,k)+\tilde G(p_0,k).
\label{k2G}
\end{eqnarray}
Substituting the above expression into Eq.~(\ref{systemofeqs2it}) and re-organizing the perturbative series we obtain
\begin{eqnarray}
\Xi_2(p_0,p) &=& p^2+q^2\int d^3  k\, G(p_0,k) \, V_\pi(k, p) \nonumber\\
&+& q^2\int d^3  k\,d^3  l\,G(p_0,k)\, V_\pi(k,l)\,G(p_0,l)\,V_\pi(l,p) +\cdots \nonumber\\
&+& \delta z +\int d^3  k\,\tilde G(p_0,k)\,V_\pi(k,p)\nonumber\\
 &+& \int d^3  k\,d^3  l\,\left[\delta z+\tilde G(p_0,k)\, V_\pi(k,l) \right]\,G(p_0,l)\,V_\pi(l,p)+\cdots .
\label{systemofeqs2itS}
\end{eqnarray}
From this equation it is easily seen that $\Xi_2(p_0,p)$ can be
written in the form
\begin{equation}
\Xi_2(p_0,p) = p^2+ q^2 \left[\Xi_1(p_0,p)-1\right] +\Psi_\pi (p_0,p) ,
\label{systemofeqs3}
\end{equation}
where the quantity $\Psi_\pi $ satisfies the equation
\begin{equation}
\Psi_\pi (p_0,p) = \xi_\pi(p_0,p)+\int d^3  k\,\Psi_\pi (p_0,k)\,G(p_0,k)\,V_\pi(k,p) \,,
\label{xxx}
\end{equation}
with
\begin{equation}
\xi_\pi(p_0,p)=\int d^3  k\,\tilde G(p_0,k)\,V_\pi(k,p)+\delta z\,,
\label{Gtilde}
\end{equation}
 or symbolically
\begin{eqnarray}
 \Psi_\pi =\xi_\pi+\Psi_\pi \,G\,V_\pi \,, \ \xi_\pi=\xi_1\,\tilde G\,V_\pi + \delta z\,.
\label{Psi2_xi_pi}
\end{eqnarray}
The $\tilde G\,V_\pi$ term in  $\xi_\pi$ contains logarithmic divergence, which can be removed
by adjusting the one-loop counter-term $\delta z$ to
\begin{equation}
\delta z=-\int d^3  k\,\tilde G(m,k)\,V_\pi(k,0)\,,
\label{V2ct}
\end{equation}
so that the quantities $\xi_\pi$, $\Psi_\pi $ and $\Xi_2$ become finite. Moreover from Eq.~(\ref{asymptotics}) it follows
that $\xi_\pi\stackrel{p\to\infty}{=}O(\ln p)$, $\Psi_\pi \stackrel{p\to\infty}{=}O(\ln p)$.

We now proceed with the renormalization of ${\cal X}$,  which in our
scheme reduces to a subtractive renormalization of $\Sigma$. The term
in
$\xi\,G\,\xi^T $ (the first diagram in the right-hand side of the
third line of Fig.~\ref{BB:fig}), which is $\delta z$-independent,
contains divergences with energy-dependent coefficients.
These divergences can be consistently subtracted using the BPHZ
prescription. Those terms in $\xi\,G\,\xi^T $ which contain $\delta
z$ linearly cancel the sub-divergences in the $\delta z$-independent
part of the two-loop diagram $\xi\,G\,V_\pi G\,\xi^T $ contained in
the $\xi\,G\,T_\pi G\,\xi^T $ part of  the quantity ${\Sigma}$ (second diagram in
the righthand side of the third line of Fig.~\ref{BB:fig}). The
overall divergence 
of the  two-loop
diagram $\xi\,G\,V_\pi G\,\xi^T $ requires an additional BPHZ
subtraction. Terms in $\xi\,G\,\xi^T $ which contain $\delta z$
quadratically cancel the two-loop sub-divergence in the $\delta
z$-independent part of the three-loop diagram $\xi\,G\,V_\pi G\,V_\pi
G\,\xi^T $ (the last explicitly shown diagram in the third line of
Fig.~\ref{BB:fig}).  In addition, all $\delta z $-dependent parts of
$\xi\,G\,\xi^T $ require an additional subtraction of overall
divergences. All other divergences appearing in the loop expansion of
$\xi\,G\,T_\pi G\,\xi^T $ are canceled automatically by contributions
of the $\delta z$ counter term. For example, the one-loop
sub-divergences of the $\delta z$-independent part of the three loop
diagram $\xi\,G\,V_\pi G\,V_\pi G\,\xi^T $ are canceled by those
expressions generated by the diagram $\xi\,G\,V_\pi G\,\xi^T $, which
are linear in $\delta z$. In the following, we provide the explicit
expressions needed to compute the quantity ${\Sigma}$ in
Eq.~(\ref{Sigma}) and define the corresponding subtractions.
It is convenient to split $\Sigma$ into three terms
\begin{eqnarray}
\Sigma=\Sigma_0+\Sigma_\pi^{\rm finite}+\Sigma_\pi^{\rm div}\,.
\label{Sigma_split}
\end{eqnarray}
The term $\Sigma_0$ contains only ``pionless'' contributions:
\begin{eqnarray}
\Sigma_0=\xi\,G\,\xi^T \arrowvert_{\delta z=0}\equiv\,\left(
\begin{array}{cc}
I_0(q) &
I_2(q)\\
I_2(q) &
I_4(q)
\end{array}
\right),
\label{sigmaX}
\end{eqnarray}
where we have introduced the integrals
\begin{equation}
\{I_0(q),I_2(q),I_4(q)\}= \int d^3 k\,\{1,\vec k^2,(\vec k^2)^2\} \,G(p_0,k)\,.
\label{defint}
\end{equation}
We subtract the infinite local (polynomial in $p_0-m$) terms from these integrals to make them finite, so that
$I_i(q)$ ($i=0,2,4$) are replaced with the subtracted $I^R_i(\mu,q)$ defined as
\begin{eqnarray}
I^R_0 (\mu,q)&=&I_0(q)-I_0(i\mu)=\frac{m^2}{8\pi^2 p_0}\left[2q\left( \,\sinh^{-1}\frac{q}{m}-i\pi\right)-\pi m\right]\nonumber\\
&+&\frac{m^2}{8\pi^2\sqrt{m^2-\mu^2}}\left[2\mu\left( \,\sin^{-1}\frac{\mu}{m}-\pi\right)+\pi m\right]\,,\nonumber\\
I^R_2 (\mu,q)&=&I_2(q)-q^2\,I_0(i\mu)-\int d^3 k\,\tilde G(p_0,k)=q^2\,I^R_0 (\mu,q)\,,\nonumber\\
I^R_4 (\mu,q)&=&I_4(q)-q^4\,I_0(i\mu)-\int d^3 k\,(k^2+q^2) \,\tilde G(p_0,k)=q^4\,I^R_0 (\mu,q)\,.
\end{eqnarray}
The subtracted integrals depend on the choice of the subtraction point
$\mu$. In principle, one has an additional
freedom in fixing finite terms
polynomial in  $p_0-m$.
However, it leads to higher-order effects,
therefore we will only study the $\mu$ dependence of obtained results.

The remaining terms of $\Sigma$ are split into the finite and divergent
parts, $\Sigma_\pi^{\rm finite}$ and $\Sigma_\pi^{\rm div}$, which are given by
\begin{eqnarray}
\Sigma_\pi^{\rm finite}&=&
\left(
\begin{array}{cc}
1 & q^2 \\
q^2 & q^4 \\
\end{array}
\right) \xi_1 \,G\,V_\pi\,G\, \Xi_1
+ \left(
\begin{array}{cc}
0 & 1 \\
1 & 2\,q^2 \\
\end{array}
\right)\xi_1 \, G\,V_\pi\,G\,\Psi_\pi \nonumber\\
&+& \left(
\begin{array}{cc}
0 & 0 \\
0 & 1 \\
\end{array}
\right)
\xi_\pi\,G\,V_\pi\,G\,\Psi_\pi \,,
\end{eqnarray}
and
\begin{eqnarray}
\Sigma_\pi^{\rm div}&=&
\left(
\begin{array}{cc}
0 & 1 \\
1 & 2\,q^2 \\
\end{array}
\right) \xi_1\, G\, \xi_\pi+ \left(
\begin{array}{cc}
0 & 0 \\
0 & 1 \\
\end{array}
\right) \Bigl\{\xi_1\,\tilde G\,V_\pi\,\tilde G\,\xi_1+2\,\delta z\, \xi_1\,\tilde G\,\xi_1
+\xi_\pi\,G\,\xi_\pi\Bigr\}\,.
\end{eqnarray}
It is straightforward to show using Eqs.~(\ref{k2G}),~(\ref{systemofeqs3}) and (\ref{Psi2_xi_pi}) that two expressions
for $\Sigma$ given by Eq.~(\ref{Sigma}) and Eq.~(\ref{Sigma_split}) are identical.
Note that the divergent part $\Sigma_\pi^{\rm div}$ contains only a
finite number of iterations of the OPE potential
(up to three loops as shown in Fig.~\ref{BB:fig}). All
nonperturbative effects due to  OPE are
included in $\Sigma_\pi^{\rm finite}$.

Again, following the BPHZ procedure, we subtract the infinite local terms
containing overall divergencies from $\Sigma_\pi^{\rm div}$ of the following form
\begin{eqnarray}
\delta\Sigma_\pi&=&
\left(
\begin{array}{cc}
0 & 1 \\
1 & 2\,q^2 \\
\end{array}
\right) \xi_1\, G_0\, \xi_\pi+ \left(
\begin{array}{cc}
0 & 0 \\
0 & 1 \\
\end{array}
\right) \Bigl\{\xi_1\,\tilde G\,V_\pi\,\tilde G\,\xi_1+2\,\delta z\, \xi_1\,\tilde G\,\xi_1
+\xi_\pi\,G_0\,\xi_\pi\Bigr\}\,,
\end{eqnarray}
where $G_0(k)\equiv G(p_0=m,k)$,
so that the full subtracted result for $\Sigma$ reads
\begin{eqnarray}
\Sigma^R&=&
\left(
\begin{array}{cc}
1 & q^2 \\
q^2 & q^4 \\
\end{array}
\right) \Bigl\{I^R_0 (\mu,q)+\xi_1 \,G\,V_\pi\,G\, \Xi_1 \Bigr\}+ \left(
\begin{array}{cc}
0 & 1 \\
1 & 2\,q^2 \\
\end{array}
\right)\Bigl\{\xi_1\, (G-G_0)\, \xi_\pi + \xi_1 \, G\,V_\pi\,G\,\Psi_\pi \Bigr\}\nonumber\\
&+& \left(
\begin{array}{cc}
0 & 0 \\
0 & 1 \\
\end{array}
\right)
\Bigl\{\xi_\pi\,(G-G_0)\,\xi_\pi + \xi_\pi\,G\,V_\pi\,G\,\Psi_\pi \Bigr\}\,.
\end{eqnarray}
The finiteness of $\Sigma^R$ can be shown using ultraviolet behavior of $V_\pi$, $\Xi_1$, $\xi_\pi$ and $\Psi_\pi $
considered above.
Performing subtractions in the spirit of chiral effective field theory we were supposed to also expand in powers
of the pion mass, which we have not done here. However, the
non-analytic dependence of the resulting subtraction terms on the pion
mass is of a higher order relative to the accuracy of our calculation.
Note also that our final perturbative result which will be discussed
in section \ref{sec3} depends on the choice of
the renormalization scheme. This dependence is also of higher order.

In final finite expressions for $\cal X$ we substitute the finite
renormalized couplings for $\tilde C$ and $C_2$\footnote{It is not
  possible to disentangle the $D$
term from the fitted value of $\tilde C^R$.}:
\begin{eqnarray}
{\cal X}^R= \left[\left({\cal C}^R\right)^{-1}-\Sigma^R\right]^{-1}\,.
\label{calX_R}
\end{eqnarray}
Note that ${\cal X}^R$ is not equal to ${\cal X}$, because not all of the divergencies can be absorbed by means of redefinition of
two available low-energy constants.
We parameterize effectively the dependence of the result on the
renormalization scheme by exploiting the
freedom to choose the subtraction point $\mu$.

\section{The choice of renormalization conditions and numerical
  results}
\setcounter{equation}{0}
\def\theequation{\arabic{section}.\arabic{equation}}
\label{sec3}

\begin{figure}
\epsfig{file=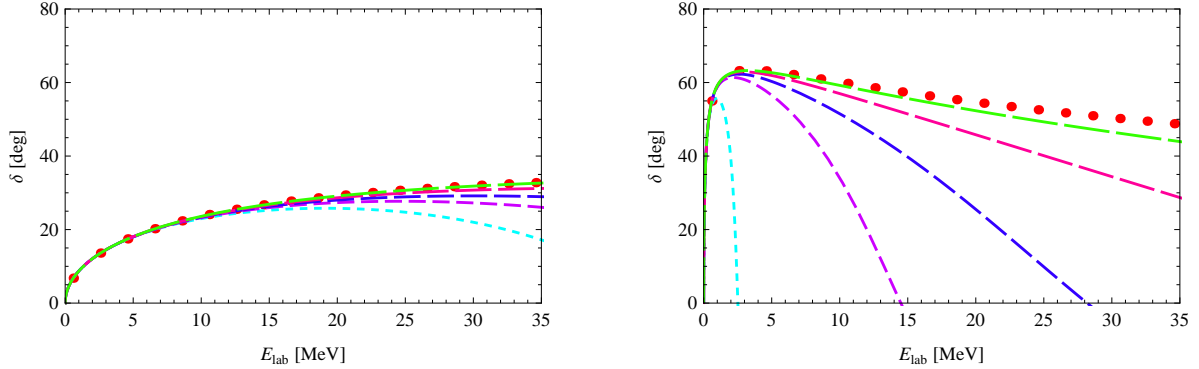, width=\textwidth}
\caption[]{\label{CID} Renormalization scale dependence of the phase
  shift in $^1S_0$ partial wave NN scattering emerging by
  non-perturbative inclusion of the NLO contact interaction in pionless
  EFT. Left panel corresponds to a natural scattering length while
  right panel shows the case of an unnaturally large scattering length. Circles
  (color online: red)
  on both panels refer to the synthetic data as described in the text
  while the dashed
  curves with increasing dash length correspond to $\mu = 1, 50, 100, 200, 400$ MeV.
}
\end{figure}

We are now in the position to specify the choice of
renormalization conditions which, in the case at hand, translates into
specifying the subtraction point $\mu$. Notice that we have already
made a specific choice for the subtractions of the integrals
$I_2 (q)$ and $I_4(q)$ in Eq.~(\ref{defint}). It is useful to recall
the key aspects of renormalization in the simple case of pionless EFT
corresponding to $g_A =0$, see e.g.~Refs.~\cite{Gegelia:1998gn,Epelbaum:2009sd},
before dealing with the more complicated pionfull approach. To be specific, consider the NN S-wave
scattering amplitude corresponding to the contact interaction
potential of Eq.~(\ref{nuCfact0}),
\begin{equation}
T_{\rm cont}=-\frac{2 \bigl\{C_2^2 m^2 \left[I(q)\, q^4-2 q^2
   I_2(q)+I_4(q)\right]+4 C_2 q^2+2 C\bigr\}}{C_2^2 m^4
   \left[I(q) I_4(q)-I_2(q){}^2\right]+4 C_2 m^2
   I_2(q)+2 C I(q) m^2-4} \,.
\label{tc}
\end{equation}
To make the following discussion more transparent, we restrict
ourselves to the leading nonrelativistic approximation so that the
above expression takes the form
\begin{equation}
T_{\rm cont}=
-\frac{C_2^2 m \left[q^4 J(q)-2 q^2 J_2(q)+J_4(q)\right]+2 C_2 q^2+C}{m
   J(q) \left[C_2^2 m J_4(q)+C\right]-\left[C_2 m J_2(q)-1\right]{}^2},
\label{TCNr}
\end{equation}
where
\begin{equation}
\{J(q),J_2(q),J_4(q)\}= \frac{2}{m} \int \frac{d^3 k}{(2 \pi)^3}\frac{\{1,\vec k^2,(\vec k^2)^2\}}{q^2- k^2+i\,\epsilon}.
\label{defintJ}
\end{equation}
We have verified via explicit calculations that the omitted
$1/m$-corrections are heavily suppressed for the problem at hand and
of no relevance for the forthcoming discussion.
Using dimensional
regularization we express $J_2(q^2)$ and $J_4(q^2)$ in terms of $J(q^2)$ as
\begin{equation}
J_2(q^2)=q^2 J(q^2), \ J_4(q^2)=q^4 J(q^2)
\label{defJ2J4}
\end{equation}
and subtract $J(q^2)$ at $q^2=-\mu^2$ obtaining
\begin{equation}
J^R(q^2)=-\frac{i\,q+\mu}{2 \pi m }.
\label{JR}
\end{equation}
Subtracted integrals $J_2^R(q^2)$ and $J_4^R(q^2)$ are obtained by
replacing $J(q^2)$ in Eq.~(\ref{defJ2J4}) by  $J^R(q^2)$
specified in Eq.~(\ref{JR}).
The renormalized amplitude is then given by Eq.~(\ref{TCNr}) with the
divergent integrals being replaced by their subtracted values and the
bare LECs  $C$ and $C_2$ being replaced by the
renormalized ones  $C^R(\mu)$ and $C_2^R(\mu)$  \cite{Gegelia:1998gn}.
Using the scattering length $a$ and effective range $r$ to determine these two
LECs, we obtain the renormalized expression for the effective range function $q \cot
\delta$ in terms of observable quantities
\begin{equation}
q \cot \delta =\frac{-a^2  r \mu \,q^2+2 a \mu
   -2}{a \left(a  r q^2 - 2 a \mu
   +2\right)}=-\frac{1}{a}+\frac{r q^2}{2}+\frac{a r^2 q^4}{4 (a \mu
   -1)}+\frac{a^2  r^3 q^6}{8 (a \mu
   -1)^2}+\ldots .
\label{pcot}
\end{equation}
Notice that the resulting expression is explicitly
$\mu$-dependent. This is because the UV divergencies emerging from
iterations of the LS equation require counter terms beyond the
truncated potential unless the $C_2$ and higher-order interactions are
treated in perturbation theory. For the natural case describing a
perturbative scenario corresponding to $a \sim
\Lambda^{-1}$,  $r \sim \Lambda^{-1}$, $\ldots$, with $\Lambda$ being
the hard scale of the order of $\Lambda \sim M_\pi$, it is appropriate
to choose the subtraction scale $\mu$ of the order of the soft scale
in the problem, i.e.~of the order of external momenta of the nucleons
$\mu \sim q \ll \Lambda$. This ensures that the values of the shape parameters $v_i$
in Eq.~(\ref{pcot}) scale with the corresponding powers of $\Lambda$,
so that the residual $\mu$-dependence in the amplitude is beyond the
accuracy of the NLO approximation.
This is visualized in the left panel of Fig.~\ref{CID}, where the
NLO pionless EFT predictions for the phase shift for the case of $a=
-M_\pi^{-1}$ fm and $r= M_\pi^{-1}$ fm are shown as a function of laboratory
energy $E_{\rm lab}$ for different choices of the subtraction scale $\mu$. The ''data'' in Fig.~\ref{CID}
correspond to the effective range approximation with all shape
coefficients set to zero.
Notice that choosing $\mu$ of the order of the hard scale also results
in a valid low-energy expansion of the effective-range function as
visualized in the figure.

On the other hand, for the unnatural case describing the non-perturbative situation of a
system being close to the unitary limit and corresponding to
very large values of the scattering length, $a\to \infty$, one obtains
from Eq.~(\ref{pcot})
the expansion
\begin{equation}
q \cot \delta =-\frac{r \mu\,q^2 }{r q^2 -2 \mu }= \frac{ r q^2}{2}+\frac{r^2 q^4}{4 \mu
   }+\frac{r^3 q^6}{8  \mu^2}+\ldots \,.
\label{pcotL}
\end{equation}
As it is clear from Eq.~(\ref{pcotL}), the generated (scheme-dependent) coefficients of
the effective range expansion will be unnaturally large if one chooses $\mu$
of the order of the soft scale in the problem. In the KSW approach, one
compensates for these large contributions by taking them into account
perturbatively and canceling against the contributions of the corresponding
higher-order contact interactions which are also assumed to be
unnaturally large (i.e.~the assumed scaling of the corresponding LECs
involves powers of the soft scale). We solve the
problem of the unnaturally large scattering length by choosing the
subtraction point $\mu$
of the order of the {\it hard scale} in the problem, $\mu \sim
\Lambda$. This  guarantees that no large contributions in the
induced coefficients of the effective range expansion are generated
and the $\mu$-dependence of the scattering amplitude is indeed beyond
the order one is working at, see Eq.~(\ref{pcot}).  This is visualized in the right panel of
Fig.~\ref{CID}. Notice that choosing $\mu \ll \Lambda$ leads to strong
distortions in the phase shifts and thus considerably restricts the
range of applicability of pionless EFT which is expected to be valid for
energies up to $E_{\rm lab}= M_\pi^2/(2 m) \sim 10.5$ MeV.
Before turning to the pionfull EFT we are actually interested in, it
is important to emphasize that the subtraction scale $\mu$ should also
not be chosen to be significantly larger than the corresponding hard
scale in the problem in order to keep $\mu$-dependent terms beyond the
accuracy of the calculation. This feature cannot be illustrated in the
considered example of pionless EFT, where taking the limit $\mu \to
\infty$ simply leads to vanishing shape parameters. In the presence of
a long-range interaction, the induced
$\mu$-dependent contributions in Eq.~(\ref{pcot}) will, in general, involve the
mass scale associated with the long-range interaction and positive powers of $\mu$. Choosing
$\mu \gg \Lambda$ will then enhance the scheme-dependent
contributions, which are nominally of a higher order, and spoil the
predictive power of a theory. An explicit example of such a
``peratization'' is considered in Ref.~\cite{Epelbaum:2009sd}.

\begin{figure}
\epsfig{file=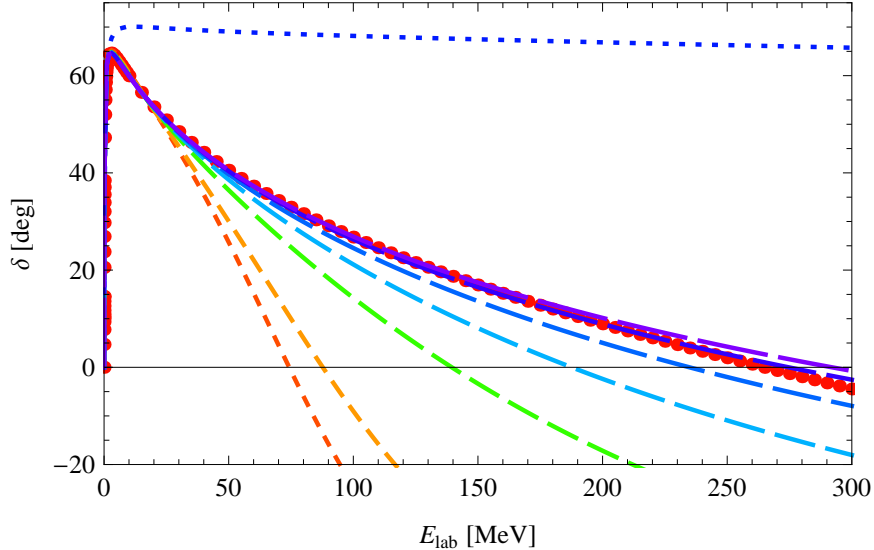, width=0.7\textwidth}
\caption[]{\label{oneloop:fig} Neutron-proton $^1S_0$ phase shifts
versus the energy in the laboratory frame. Circles (red)
correspond to the Nijmegen PWA \cite{Stoks:1993tb}. Dotted line represents the LO result. Curves with different dashing (colors) correspond to
non-perturbative inclusion of the NLO contact
interaction potential for $\mu = 50, 100, 300, 500, 700, 850$ and  $900$
MeV respectively.}
\end{figure}

After these introductory remarks, we are now in the position to
present our results for the $^1S_0$ phase shift at NLO in chiral EFT.
We employ the exact isospin symmetry as appropriate at LO and
use the following values for the LECs entering the OPE potential
\beq
\label{LECs}
M_\pi = 138 \mbox{ MeV}, \quad \quad
F_\pi = 92.4 \mbox{ MeV}, \quad \quad
g_A = 1.267\,.
\eeq
The numerical value of the renormalized LECs $\tilde C^R (\mu )$ and $C_2^R
(\mu)$ are determined from a fit to the neutron-proton  $^1S_0$ phase
shift of the Nijmegen partial-wave analysis (PWA) \cite{Stoks:1993tb}
in the energy range of $0\ldots 50$ MeV
for several choices of the subtraction point $\mu$ as discussed above.
The resulting phase shifts for different choices of $\mu$ are plotted in Fig.~\ref{oneloop:fig}.
As expected and explained at the beginning of this section, we do
observe some residual $\mu$-dependence of the predicted phase shifts
which gets strongly enhanced if one chooses $\mu$ of the order or
smaller than  $M_\pi$. On the
other hand, for the appropriate choice of the subtraction scale
$\mu \sim \Lambda$,  where the hard scale $\Lambda$ can be
realistically estimated to be of the order of
$\Lambda \sim 600$-$700$ MeV\footnote{These values are of the
  order of the masses of the sigma and rho mesons which
  phenomenologically are known to yield the most important short-range
 contribution to the nucleon-nucleon potential \cite{Epelbaum:2001fm}. This
 estimation also agrees well with the findings of chiral EFT
 calculations utilizing a finite cutoff \cite{Epelbaum:2014efa,Epelbaum:2014sza}. A deeper discussion on the
 breakdown scale of nuclear chiral EFT can be found in
 Ref.~\cite{Lepage:1997cs}.}, the dependence on $\mu$ appears to be
moderate, and the predicted energy dependence of the phase shift shows a
good agreement with the Nijmegen PWA. Choosing $\mu
\sim \Lambda$,  the observed $\mu$-dependence
of the phase shift is considerably smaller than the
difference between the LO and NLO results and can serve as an
estimation of the size of corrections beyond NLO, i.e.~it defines the
lower bound for the theoretical uncertainty
of our calculation. It is especially comforting to see that the spread of
predictions for the subtraction point chosen in the range of $\mu =
500\ldots 900$ MeV matches very well the estimated theoretical accuracy at
NLO in calculations based on a finite cutoff, see Fig.~9 of Ref.~\cite{Epelbaum:2014efa}

\begin{figure}
\epsfig{file=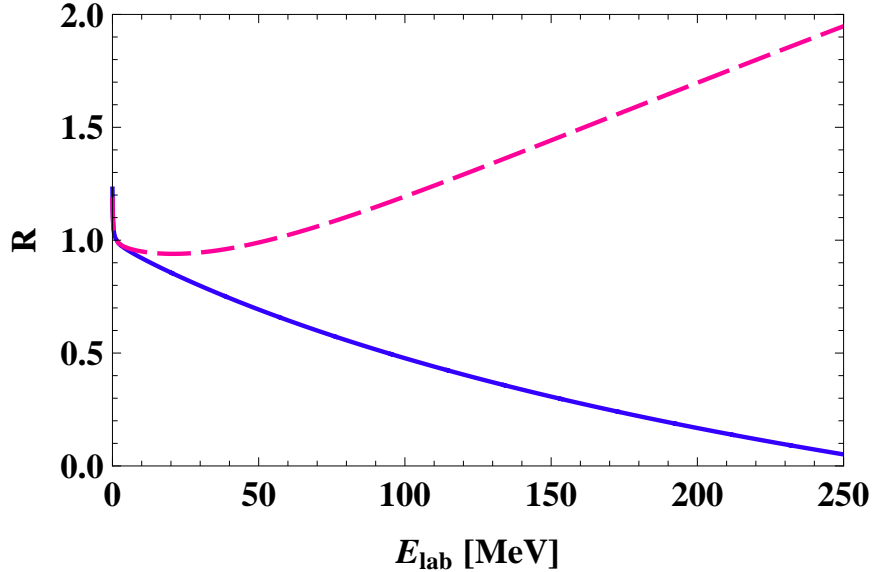, width=0.7\textwidth}
\caption[]{\label{oneloopP:fig} The ratio $R$ defined in Eq.~(\ref{RatioR})
  versus the energy in the laboratory frame. The solid (dashed) line shows the
  result based on $T^{\rm NLO}$ corresponding to the nonperturbative (perturbative)
  inclusion of the subleading contact interaction.  In both cases, the
  subtraction scale is set to $\mu = 850$ MeV.
}
\end{figure}

It is also interesting to address the question of perturbativeness of
the subleading short-range interaction within our scheme. Given that
calculating phase shifts always relies on some kind of unitarization
procedure, it is more appropriate to address this issue by looking at
the scattering amplitude directly. To be specific, consider the ratio
$R \big(E_{\rm lab} \big)$ defined as
\begin{equation}
\label{RatioR}
R \big(E_{\rm lab} \big) = \frac{\big| T^{\rm NLO} \big(E_{\rm lab}
  \big) \big|}{\big| T^{\rm LO} \big(E_{\rm lab}
  \big) \big|}\,,
\end{equation}
where  $T^{\rm LO}$ and $T^{\rm NLO}$ denote the $T$-matrix calculated
at LO and up to NLO, respectively. In Fig.~\ref{oneloopP:fig}  we show by
the solid line the quantity $R$ based on the nonperturbative inclusion
of the subleading contact interaction as described above and
corresponding to the choice of $\mu = 850$ MeV.\footnote{Note that the ratio R is not equal to 1 at threshold, because in our
subtraction scheme diagrams containing NLO contact interactions
but no overall divergencies are not subtracted.} Notice that the
subleading contribution to the amplitude becomes comparable in size
with the leading one at higher energies, the feature that could have
been expected by looking at the LO prediction for the $^1$S$_0$ phase
shift.  We also plot in this figure the ratio $R$ resulting from the
inclusion of the subleading contact interaction in first-order
perturbation theory  for the same choice of $\mu$ (and using the same values for the renormalized
low-energy constants as determined in the nonperturbative calculation).
This shows clearly that it is advantageous to include the subleading
contact interaction nonperturbatively within the employed framework
for energies of about $E_{\rm lab} \sim 50$ MeV and higher.

Last but not least, we also give the predictions for the
coefficients in the effective range expansion which may be regarded as
low-energy theorems (LETs), see \cite{Cohen:1998jr,Epelbaum:2009sd} for more details.
In table \ref{let_1s0}, the LETs in the  KSW and Weinberg
approaches are confronted with the results of the Nijmegen PWA
for the $^1S_0$ partial wave, respectively.
\begin{table}[t]
\caption{Predictions for the coefficients in the effective
  range expansion of the $^1S_0$ phase shifts (low-energy theorems)
 at LO and NLO in the modified Weinberg approach in comparison with
 the NLO KSW predictions of Ref.~\cite{Cohen:1998jr} and empirical
 numbers extracted from the Nijmegen PWA \cite{Stoks:1994wp,PavonValderrama:2005ku}.
For the NLO Weinberg results, we show the predictions corresponding to the
variation of the subtraction point in the
range of $\mu = 500 \ldots 900$ MeV.  The errors quoted for the LO
predictions refer to the uncertainty in the numerical extraction of
the coefficients \cite{Epelbaum:2012ua}.
\label{let_1s0}}
\smallskip
\begin{tabular*}{\textwidth}{@{\extracolsep{\fill}}lrrrrr}
\hline
\hline
\noalign{\smallskip}
& $a$  [fm] & $r$  [fm] & $v_2$  [fm$^3$]& $v_3$  [fm$^5$]  & $v_4$  [fm$^7$] \smallskip
 \\
\hline
\hline
LO, Ref.~\cite{Epelbaum:2012ua} &  fit  & $1.50$  &
$-1.9$ &$8.6(8)$  &$-37(10)$
\\
NLO, nonperturb.~$C_2$ &  fit  & fit  &
$-0.55 \ldots -0.61 $ &$ 5.1 \ldots 5.5 $  &$ -29.6 \ldots -30.8 $
\\
NLO, perturbative $C_2$  &  fit  & fit  &
$ -0.51 \ldots -0.57 $ &$ 4.5 \ldots 4.7 $  &$ -28.8 \ldots -29.8 $
\\
\hline
NLO KSW, Ref.~\cite{Cohen:1998jr}  & fit   & fit  & $-3.3$  & $18$ & $-108$
\\\hline
Nijmegen PWA & $-23.7$  & $2.67$  & $-0.5$  & $4.0$ & $-20$
\\[2pt]
\hline \hline
\end{tabular*}
\end{table}
We observe a clear improvement in the reproduction of the LETs when
going from LO to NLO. Notice that the nonperturbative treatment of the
subleading contact interaction appears to have minor effect for the LETs. It
should, however, be emphasized that the extraction of the coefficients
in the effective range expansion requires performing a unitarization
of the amplitude which provides a partial resummation of
$C_2$-contributions.

\section{Summary and conclusions}
\def\theequation{\arabic{section}.\arabic{equation}}
\label{sec4}

In this paper we have considered nucleon-nucleon
scattering in the $^1S_0$ partial wave within the modified Weinberg
approach. The integral equation based on the leading-order potential,
which consists of the momentum- and energy-independent contact
interaction and the OPE potential, is renormalizable and was studied
in Ref.~\cite{Epelbaum:2012ua}.
The observed large discrepancy between the LO EFT results and the
$^1$S$_0$ phase shift of the Nijmegen PWA, which starts already at
rather low energies, indicates that at least some parts
of the higher-order contributions to the effective potential need to
be included nonperturbatively.
Here we assumed that only the short range part of the NLO potential
(in standard Weinberg power counting) needs to be treated
non-perturbatively. It involves only the contact
interaction terms quadratic in momenta and the pion mass. This makes
it possible to perform the subtractive renormalization explicitly in
non-perturbative expressions.

The pertinent results of our study can be summarized as follows:
\begin{itemize}
\item
We have carried out subtractive renormalization of the scattering
amplitude based on the potential involving OPE as well as the leading and
subleading contact interactions using the framework of Ref.~\cite{Epelbaum:2012ua}
and without relying on perturbation theory.
\item
The resulting renormalized integral equations for the scattering
amplitude have been solved
numerically and the values of the renormalized low-energy constants
$\tilde C^R (\mu)$ and $C_2^R (\mu)$ were determined from a fit to phase
shifts of the Nijmegen PWA for different choices of the subtraction
point $\mu$.
\item
We discussed the issue of the proper choice of renormalization
conditions in our scheme and have argued that the observed large value
of the scattering length requires choosing the scale $\mu$, which corresponds to the renormalization of the LO contact
interaction, of the order of the {\it hard} scale in the problem.
\item
The resulting preditions for the energy dependence of the $^1$S$_0$
phase shift are in a good agreement with the Nijmegen PWA. Moreover,
the observed dependence of the phase shifts on the subtraction point $\mu$ chosen in the
range of $\mu = 500 \ldots 900$ MeV agrees well with the theoretical
accuracy at NLO estimated in calculations of Refs.~\cite{Epelbaum:2014efa,Epelbaum:2014sza} based on the
standard non-relativistic framework with a finite cutoff.
\item
We have also addressed perturbativeness of the subleading contact
interaction within our scheme. We found that it is advantageous to treat the subleading
contact interaction nonperturbatively at energies of about $E_{\rm lab} \sim 50$ MeV and higher.
\item
Finally, we have looked at the low-energy theorems for the
coefficients in the effective range expansion and found a clear
improvement when going from LO to NLO.
\end{itemize}
The results of our work open the way to perform higher-order
calculations within the modified Weinberg approach proposed in
Ref.~\cite{Epelbaum:2012ua}. As a next step, the role of the two-pion
exchange potential needs to be investigated and the extension to other
partial waves has to be performed. Work along these lines is
in progress.

\section*{Acknowledgments}
\label{sec_app}
This work is supported by the Deutsche Forschungsgemeinschaft
(SFB/TR 16, ``Subnuclear  Structure of Matter'' and GE 2218/2-1),
by the European Community Research Infrastructure Integrating Activity
``Study of Strongly Interacting Matter''
(acronym HadronPhysics3, Grant A\-gree\-ment n.~283286) under the 7th
Framework Programme of the EU, by
the European Research Council (acronym NuclearEFT, ERC-2010-StG 259218)
and by the Georgian Shota Rustaveli National Science Foundation (grant 11/31).


\end{document}